\documentclass[journal,transmag]{IEEEtran}
\usepackage{cite}
\usepackage{graphicx}
\usepackage{amssymb,amsmath}
\usepackage{makecell}
\usepackage{xcolor}
\usepackage{soul}
\ifCLASSINFOpdf

\else
\fi

\begin{document}
\title{Room-temperature Ferroelectric Control of 2D Layered Magnetism}
\author{\IEEEauthorblockN{Yingying Wu\IEEEauthorrefmark{1}, 
Zden\u{e}k Sofer\IEEEauthorrefmark{2}, 
Muthumalai Karuppasamy\IEEEauthorrefmark{2}, and
Wei Wang\IEEEauthorrefmark{3}}
\IEEEauthorblockA{\IEEEauthorrefmark{1}Department of Electrical and Computer Engineering,
University of Florida, Gainesville, FL 32611, USA}
\IEEEauthorblockA{\IEEEauthorrefmark{2}Department of Inorganic Chemistry, University of Chemistry and Technology Prague, \\Technick\'a 5, 166 28 Prague 6, Czech Republic}
\IEEEauthorblockA{\IEEEauthorrefmark{3}Department of Quantum Matter Physics, University of Geneva, \\24 Quai Ernest Ansermet, CH-1211, Geneva, Switzerland}

\thanks{
Corresponding author: Y. Wu (email: yingyingwu@ufl.edu).}}

\IEEEtitleabstractindextext{%
\begin{abstract}
 Electrical tuning of magnetism is crucial for developing fast, compact, ultra-low power electronic devices. Multiferroics offer significant potential due to their ability to control magnetism via an electric field through magnetoelectric coupling, especially in layered ferroelectric/ferromagnet heterostructures. A key challenge is achieving reversible and stable switching between distinct magnetic states using a voltage control. In this work, we present ferroelectric tuning of room-temperature magnetism in a 2D layered ferromagnet. The energy-efficient control consumes less than 1 fJ per operation which is normally in the order of several 10$^{-3}$ fJ, resulting in a $\sim$ 43\% change in magnetization. This tunable multiferroic interface and associated devices provide promising opportunities for next-generation reconfigurable communication systems, spintronics, sensors and memories.     
\end{abstract}

}

\maketitle

\IEEEdisplaynontitleabstractindextext
\IEEEpeerreviewmaketitle

\section{Introduction}
\begin{figure}
\centering
	\includegraphics[width=3.4in]{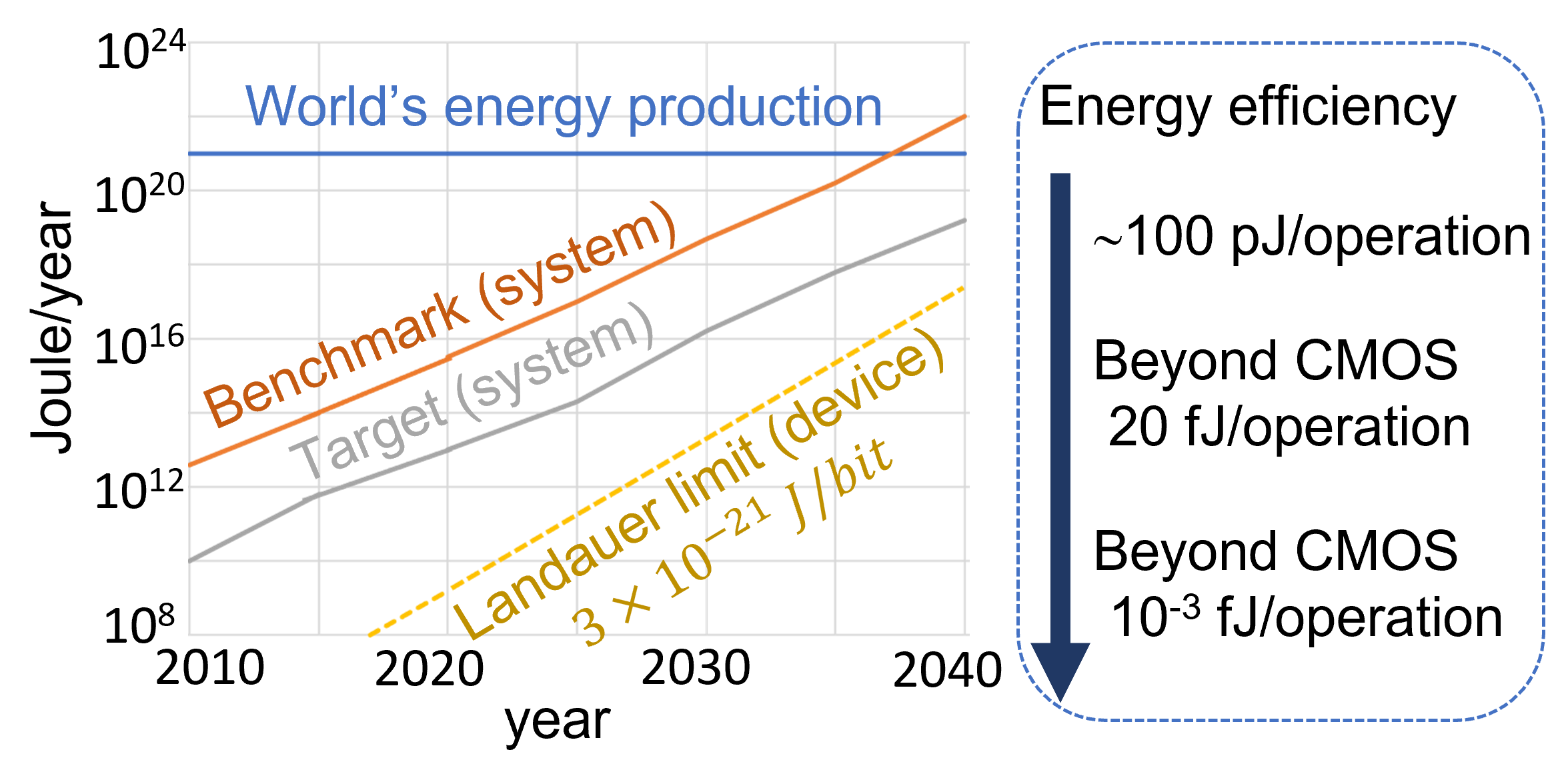}
 \vspace{-10pt}
	\caption{Energy consumption trend in computing, potentially due to AI and internet of things. Fundamental science is required to below fJ/operation energy scale. }
	\label{energy}
\end{figure}

\IEEEPARstart{E}{lectric} field control of spin and magnetic states has become a key focus over the past decade due to the demand for faster, more compact, and more energy-efficient electronic devices\cite{zhang20232d,hu2019perspective,ramesh2021electric,chen2016probing,wu2016negative,zhong2024integrating,hendriks2024electric,wang2024manipulation}. Traditional approaches using magnetic fields and electric currents are more energy-intensive compared to electric fields, which can potentially reduce energy consumption by several orders of magnitude. Considering the current energy consumption trend driven by artificial intelligence (AI), several AI companies may consume as much power as an entire country, highlighting the urgent need for energy-efficient hardware to reduce power usage in AI tasks (Fig. \ref{energy}). Aligning with global energy production limits, we urgently need to reduce the energy consumption of electronic systems. Previously, energy efficiency was at 100 pJ per operation. Current developments have reduced this value to several tens or hundreds of fJ per operation. Our future goal is to achieve $10^{-3}$ fJ per operation. Additionally, as data storage devices shrink, the local magnetic fields needed to write a single bit can interfere with neighboring bits, causing data instability. The solution lies in developing new materials and functionalities to integrate into non-volatile, low-power electronic devices. Multiferroics, which can change the magnetic state by applying an electric field through magnetoelectric (ME) coupling, offer a promising solution. Progress has been made in electric tuning of ferromagnetic resonance\cite{min2023reduced}, magnetoresistance\cite{chen2019giant}, and exchange bias\cite{borisov2005magnetoelectric} in multiferroic heterostructures.

Among low-dimensional systems, freestanding two-dimensional (2D) materials\cite{leger2024machine,zhong2024integrating,han2018investigation,wu2019induced} exhibit weak van der Waals (vdW) interlayer interactions, making them ideal for studying the interplay of various electronic and magnetic phenomena. 2D vdW magnets maintain long-range magnetic order with atomically thin layers, with a thickness down to $\sim$0.8 nm, facilitating easy electric control. Additionally, their weak interlayer interactions enable the formation of high-quality, sharp interfaces in heterostructures. These heterostructures can be tailored to incorporate specific physical properties. For example, interfacing with transition metal dichalcogenides is utilized when strong spin-orbit coupling is required\cite{wu2020neel,wu2019induced}. Large-scale exfoliation method with a lateral size of millimeter to centimeter was also developed for integration of this materials\cite{velicky2018mechanism,huang2020universal}. Electric control of 2D magnetism has been reported, predominantly at cryogenic temperatures. For instance, extensive studies have focused on electrically manipulating CrI$_3$ layers \cite{huang2017layer,huang2018electrical, wang2018electric}, despite its instability in air and low Curie temperature ($T_\textrm{C}$) around 45 K. Ionic liquid gating has shown promise in increasing the Curie temperature of Fe$_3$GeTe$_2$ from $\sim$ 200 K to $\sim$ 300 K \cite{deng2018gate}. Similarly, ME coupling can tune carrier density over a wide range up to 10$^{14}$ cm$^{-2}$ \cite{toprasertpong2022strong, zhang2010tuning}, yet its application in ferroelectric tuning of 2D magnetism remains underexplored.
Recently, Fe$_3$GaTe$_2$ (FGaT) has emerged as a room-temperature ferromagnet, even in monolayers, boasting $T_\textrm{C}$ up to 380 K for multilayers \cite{zhang2022above}. It exhibits 2D magnetism with a high Curie temperature and robust perpendicular magnetic anisotropy.

In our work, we demonstrate room-temperature ferroelectric control of magnetism using transport measurements. Our approach achieves energy-efficient operation below 1 fJ and observes $\sim$ 43\% magnetization tuning in a 20-nm thick 2D magnet FGaT, paving the way for new applications in low-power electronic devices.   

\section{Heterostructure Design}
To realize room-temperature ferroelectric tuning, we need both room-temperature ferroelectrics and ferromagnet. FGaT is chosen for its high $T_\textrm{C}$ , which is ideal for room-temperature devices considering possible heated environment up to 380 K in industry. CuInP$_2$S$_6$ (CIPS) is selected as the ferroelectric layer for its ferroelectricity (3 $\mu$C/cm$^2$) up to 320 K and a relatively large band gap (2.62 eV\cite{ma2020high}). These two materials both have layered structure, leading to high-quality interface thanks to the weak interlayer interaction. The interaction at their interface is also vdW-type, the same interaction in its pristine layers. 

\begin{figure}
\centering
	\includegraphics[width=3in]{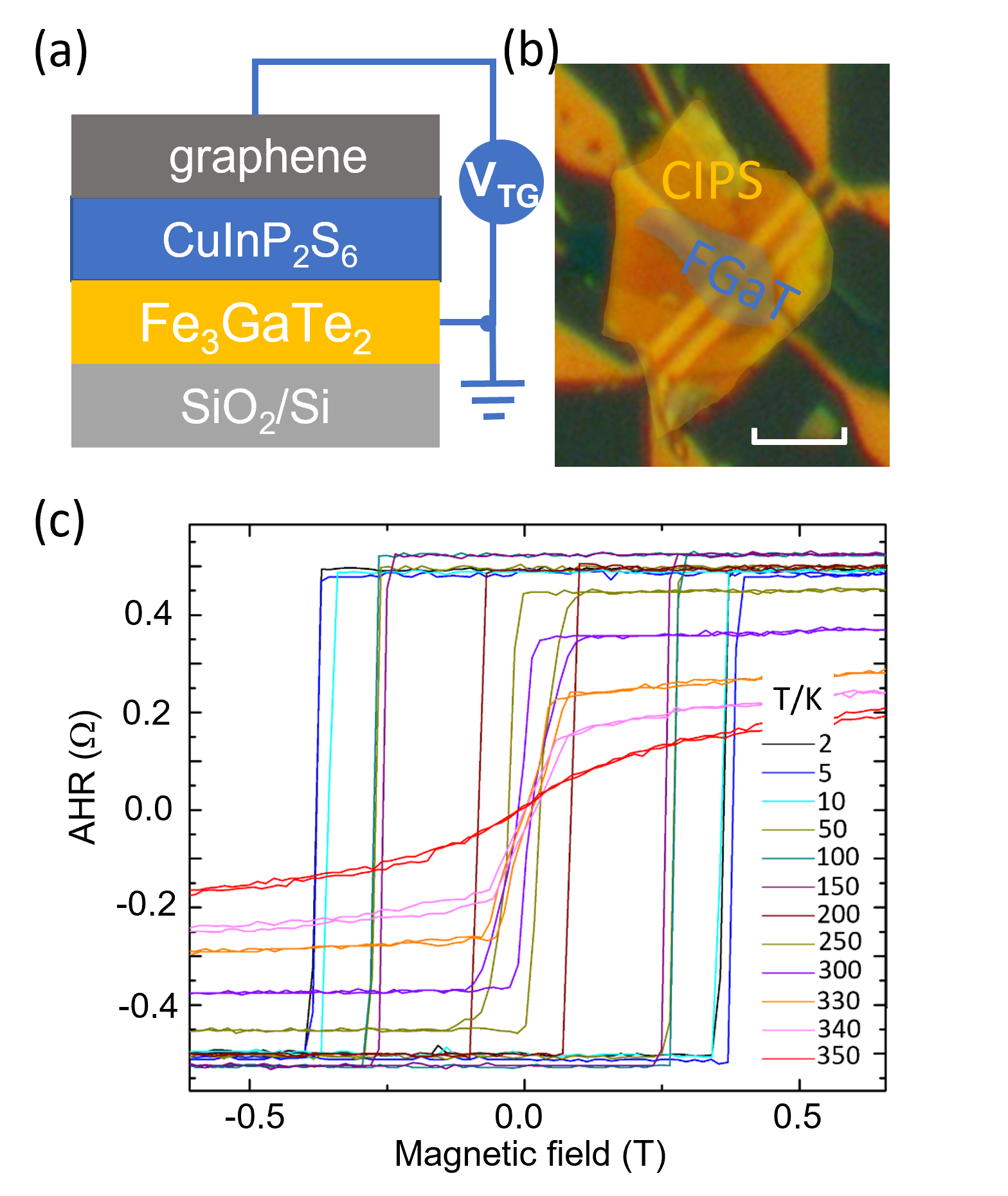}
 \vspace{-10pt}
	\caption{2D magnetic device structure and test. (a) Schematic structure for electric gating through ferroelectric layer. (b) False-color optical image of device, with a scale bar: 5 $\mu$m. (c) Anomalous Hall effect exists above room temperature.}
	\label{device}
\end{figure}
The device structure is illustrated in Figs. \ref{device}a-b, with a few-layer graphene serving as the metal contact. Mechanical exfoliation with scotch tape was used for thin flakes, followed by pick up-transfer technique to assemble the heterostructure. After fabrication, the device exhibits a lateral size of a few micrometers. In our exfoliation process, we can also obtain large-size FGaT with a lateral size of 1.6 mm, promising for future integration into large-scale device arrays. But in this experiment, demonstration is made on a microscale device. The gate voltage applied through the graphene reaches the ferroelectric layer first, altering the electric polarization in CIPS. This, in turn, induces ME coupling at the ferroelectric and ferromagnet interface, leading to a change in magnetization. ME effect can be explained using varied mechanisms, like carrier density tuning\cite{deng2018gate}, multiferroic coupling\cite{chen2021reversal,song2017recent} and layer-resolved magnetism\cite{gong2018electrically}. In the Fe$_3$GeTe$_2$ case\cite{deng2018gate}, ionic liquid gating was used to tune the ferromagnetism and increase the Curie temperature. It was explained through occupying of 3d orbitals from larger carrier density. Our case of Fe$_3$GaTe$_2$ has a carrier density modification in the order of 10$^{13}$ cm$^{-2}$ from the normal Hall effect. Additionally, the strain exerted is very difficult to quantify, but based on previous work\cite{schlom2007strain}, it likely results in only a few percent change in the lattice parameters. Thus carrier density plus strain would be the most possible mechanism for the ME coupling. Thickness dependence studies will offer a better understanding of the ME coupling at room temperature, as future work.   

\section{Results}
\subsection{Above RT ferromagnetism}
The anomalous Hall resistance (AHR) in the heterostructure was measured using lock-in technique, with a Hall-bar configuration of bottom electrode. This was done in a PPMS quantum design system with a temperature range of 2-400 K and a magnetic field up to $\pm$ 9 T. During the measurement, the temperature was varied from 2 K to 350 K, with a magnetic field applied within $\pm$ 1 T. From Fig. \ref{device}c, the hysteresis loop persists up to $\sim$340 K, close to the $T_\textrm{C}$ in 20 nm-thick FGaT layer. The coercive field of this device is below $\pm$ 0.5 T, with a AHR value around 1 $\Omega$. Both AHR and coercive fields show dependence on the temperature, indicating the magnetic anisotropy is effected by thermal fluctuations.   

\subsection{Ferroelectric tuning of 2D magnetism}
Ferroelectric gating was applied through the graphene layer, with gate voltages ($V_\textrm{TG}$) ranging between $\pm$ 15 V. This maximum value was determined by the leakage current vertically passing through the CIPS layer, which reaches up to 50 nA at $V_\textrm{TG} = \pm$ 15 V. To prevent potential breakdown of the ferroelectric layer, the applied voltage was kept within this range. Considering the voltage pulse in the nanosecond range, the energy for one operation is estimated to be below 1 fJ, typically in the order of 10$^{-3}$ fJ, demonstrating energy efficiency.

\begin{figure}
\centering
	\includegraphics[width=3.4in]{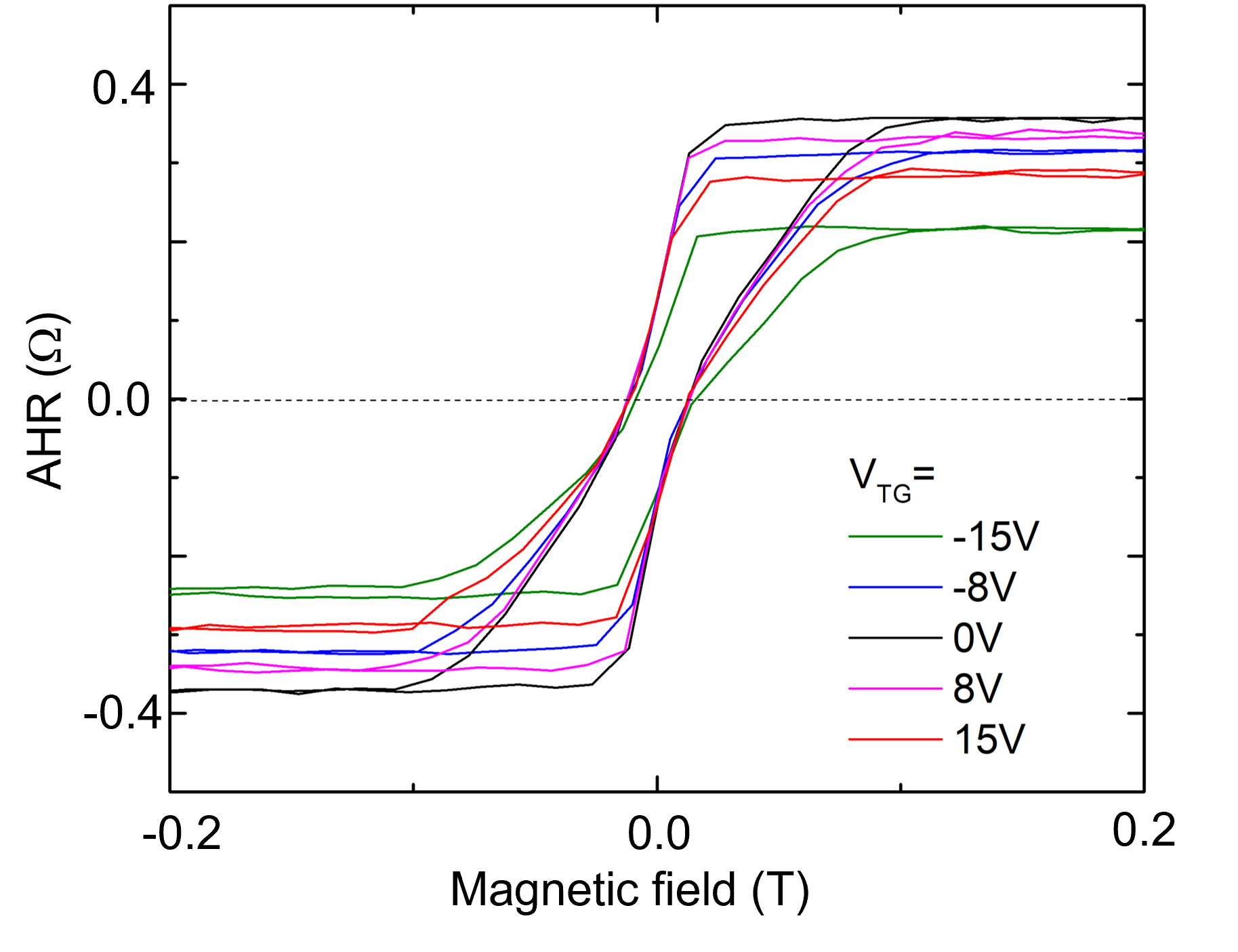}
  \vspace{-10pt}
	\caption{Ferroelectric tuning of anomalous Hall resistance at room temperature.}
	\label{tuning}
\end{figure}

At room temperature shown in Fig. \ref{tuning}, the AHR can be tuned by varying $V_\textrm{TG}$, with a maximum value at $V_\textrm{TG} = 0$ V. When $V_\textrm{TG} = -15$ V, the AHR is significantly reduced. The coercive fields are slightly adjusted by the ferroelectric gating. The AHR of a ferromagnet is known to be proportional to the saturation magnetization $M_\textrm{s}$, i.e., $\rho_\textrm{xy}^{AH} = R_\textrm{s}M_\textrm{s}$, where $R_\textrm{s}$ is the anomalous Hall coefficient. It should be noted that this relationship may not apply uniformly across different material systems. The anomalous Hall effect (AHE) in ferromagnets can originate from either intrinsic mechanisms\cite{fang2003anomalous}, related to the electronic band structure or Berry phase\cite{zheng2021berry}, or extrinsic mechanisms, due to the scattering of charges by impurities or defects with large spin-orbit coupling via skew-scattering or side-jump mechanisms\cite{nagaosa2010anomalous}. Both mechanisms may play a role in the present case.      

\begin{figure}
\centering
	\includegraphics[width=3.2in]{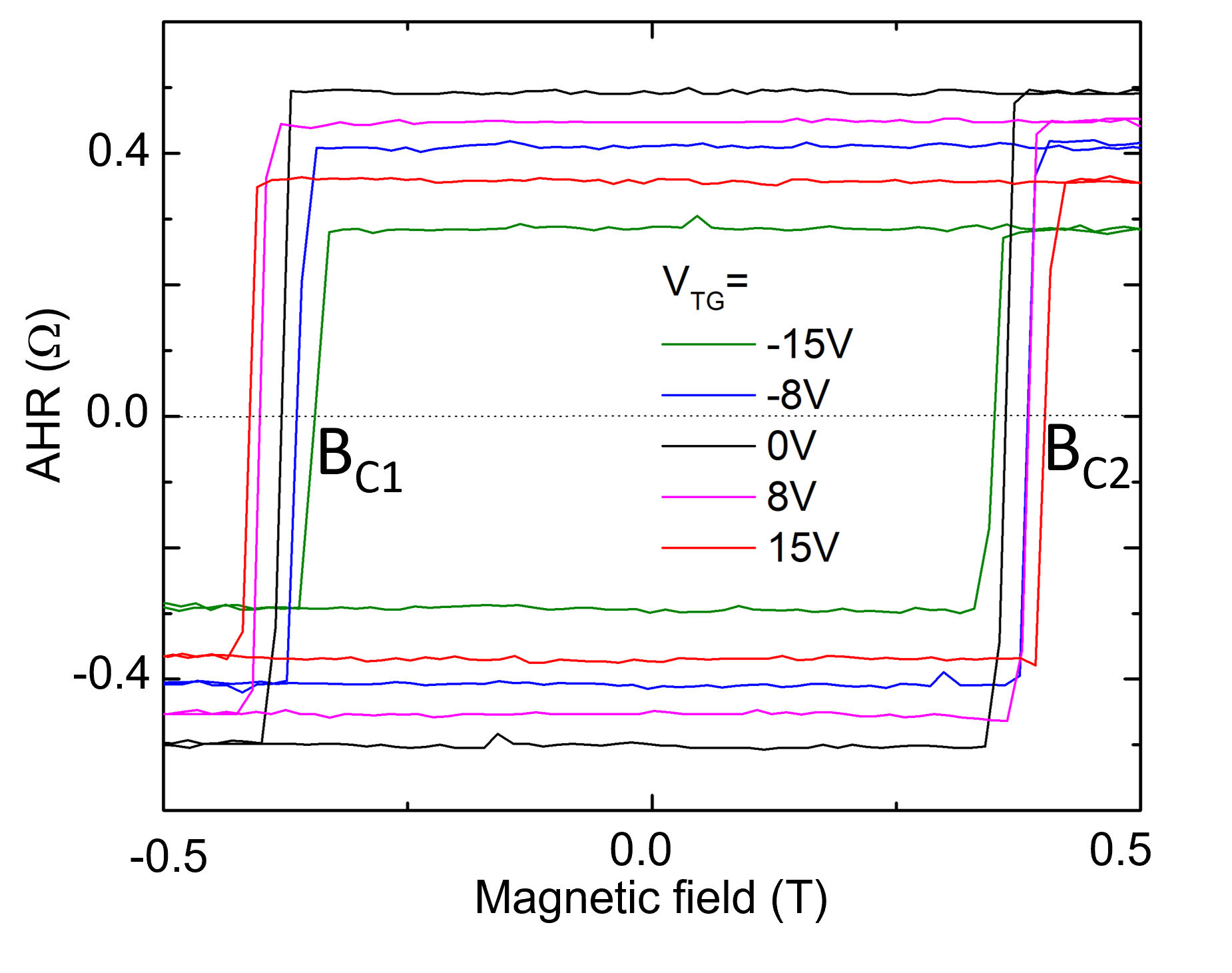}
  \vspace{-10pt}
	\caption{Ferroelectric tuning of anomalous Hall resistance at 2 K.}
	\label{2Ktuning}
\end{figure}

At 2 K, which is lowest temperature for measurement in this work, the ferroelectric gating not only tune the AHR but also the coercive fields (Fig. \ref{2Ktuning}). Similar to 300 K, AHR reaches maximum value at 0 V top gate voltage. Different from 300 K case, the coercive field shows large modulation when gate voltage changes. We can define two coercive fields as the intersection of the magnetic hysteresis loop with a y-axis value of 0, i.e. $B_\textrm{C1}$ and $B_\textrm{C2}$. Detailed analysis will be carried out in Section. \ref{Sec:exchange}.

\subsection{Magnetization tuning}
\begin{figure}
\centering
\includegraphics[width=3.2in]{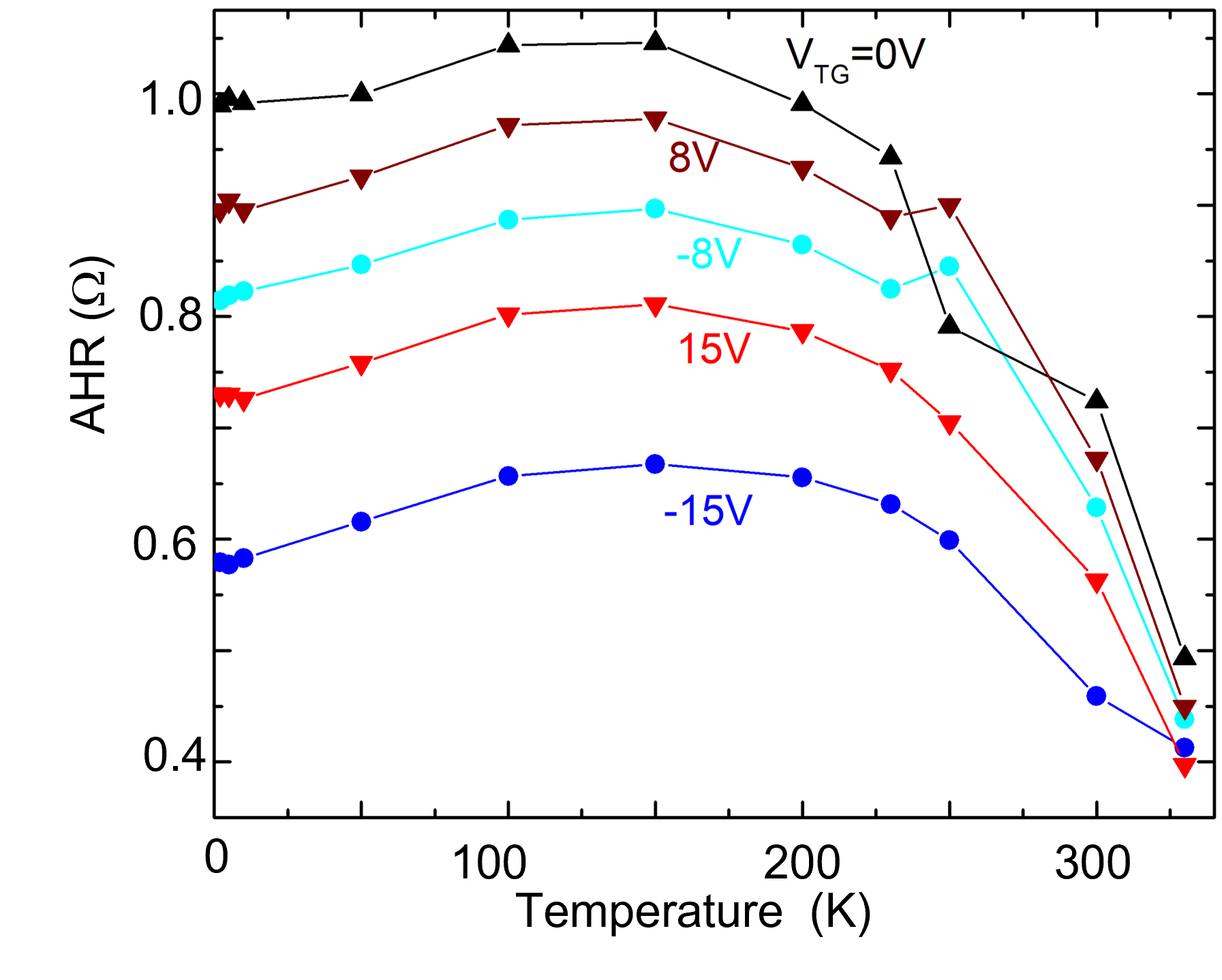}
  \vspace{-10pt}
	\caption{$\sim$ 43\% magnetization tuned by ferroelectric gating.}
	\label{percent}
\end{figure}

This AHR is proportional to the FGaT magnetization. We measured the AHR at varied temperatures from 2 K to 350 K, with gate voltages applied. The extracted AHR shows both temperature and gate voltage dependence (Fig. \ref{percent}). In Fig. \ref{percent}, we extract the AHR and show its dependence on both the temperature and gate voltages. $V_\textrm{TG}= \pm\,15$ V shows the large modification to the resistance. Among all the temperatures, a large modification of magnetization was obtained with a ratio of $\sim$ 43\%. This tuning was realized in a 20 nm-thick FGaT-based device, where a even larger tuning ratio is expected with reduced FGaT thickness, owing to expected lower carrier density ($\sim 10^{13}$ cm$^{-2}$). While for a ferroelectric field-effect transistor, the charge density at the interface could be tuned in the order of $10^{14}$ cm$^{-2}$\cite{toprasertpong2022strong,zhang2010tuning}. Thus, it is feasible for future work to tune 100\% of nanometer-thick FGaT magnetization.

\subsection{Exchange bias dependence on temperature and gate voltage}\label{Sec:exchange}
\begin{figure}
\centering
	\includegraphics[width=3.2in]{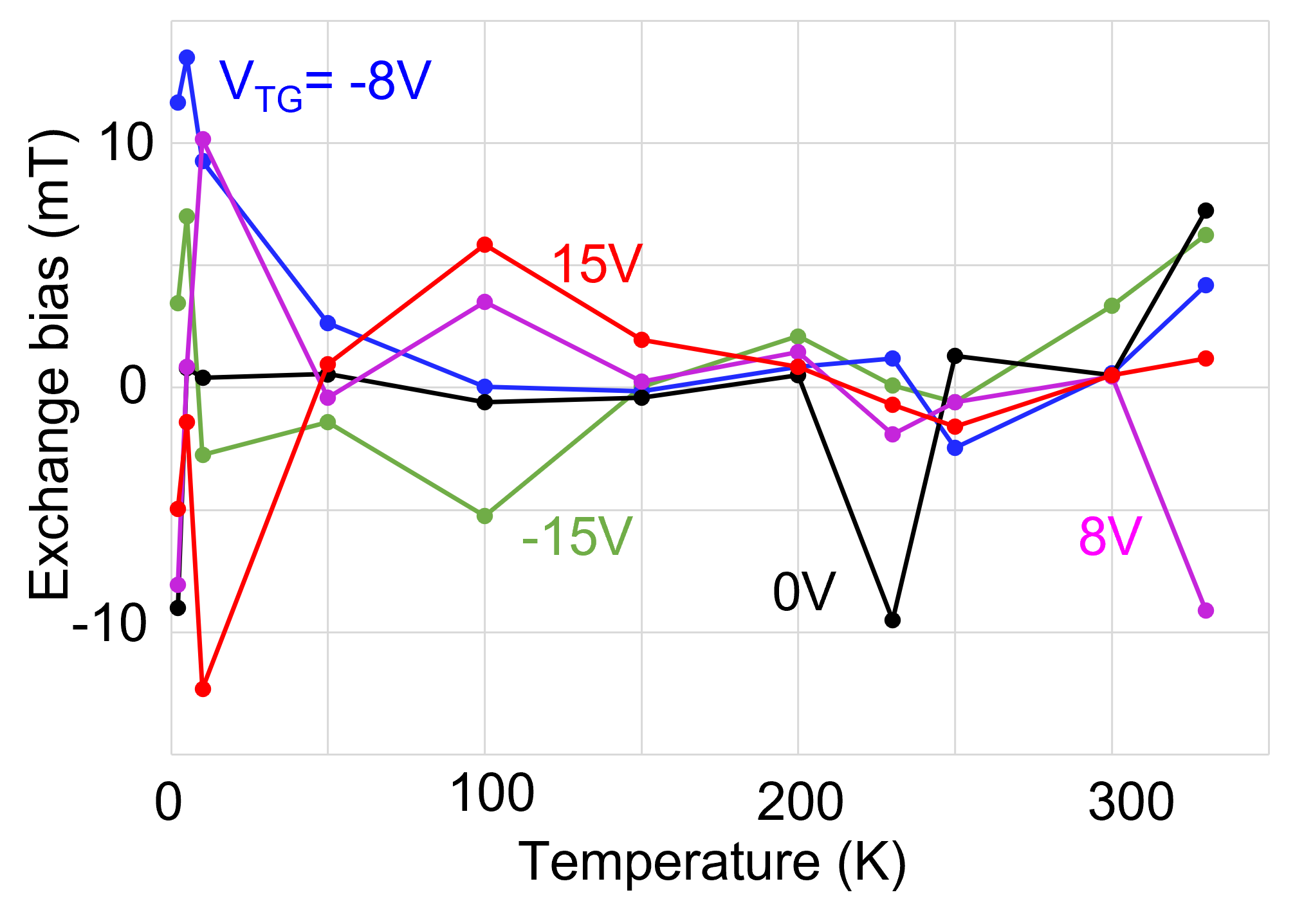}
  \vspace{-10pt}
	\caption{Exchange bias dependence on gate voltages and temperature.}
	\label{exchange}
\end{figure}

Exchange bias occurs in bilayers (or multilayers) of magnetic materials where the magnetization of a ferromagnetic film is pinned by that of an antiferromagnet\cite{gweon2021exchange,wu2022manipulating}. It has become an integral part of modern magnetism, as it provides a well-defined principal direction of spin polarization for spintronic devices. In the case of Fe$_3$GeTe$_2$, exchange bias was reported at the interface of this material with antiferromagnets like CrSe\cite{wu2022manipulating}, FePSe$_3$\cite{huang2023manipulating}, and oxide-Fe$_3$GeTe$_2$\cite{wu2022manipulating,gweon2021exchange}.

The exchange bias was defined as $B_\textrm{ex}=\frac{B_\textrm{C1}+B_\textrm{C2}}{2}$. The shift of the loop center from 0 T field indicates the potential exchange bias in this system. In our setup, the exchange bias is largest at 5 K with a gate voltage $V_\textrm{TG}$= -8 V, with a value of 13.8 mT (Fig. \ref{exchange}). At room temperature of 300 K, the exchange bias is very small. When the temperature is increased to 330 K, the exchange bias can be around several mT. One possible origin for this exchange bias is pinning of soft magnetization to hard magnetization layers. As in the case of Cr$_2$Ge$_2$Te$_6$/Fe$_3$GeTe$_2$ structures\cite{wu2022van}, even though both are ferromagnets, a small exchange bias can be observed in the order of a few mT. Compared to Cr$_2$Ge$_2$Te$_6$, Fe$_3$GeTe$_2$ shows much stronger magnetic anisotropy and a larger coercive field, which can serve as the hard magnetization layer in the exchange bias, functioning similarly to an antiferromagnetic layer.

In this work, ferroelectric tuning affects only the surface layers to a depth of a few nanometers, leaving the underlying FGaT layers largely untuned. These deeper layers retain their strong magnetization, acting as the hard magnetization layer. This selective tuning could be a potential origin of the exchange bias observed. Regarding the detailed mechanism, it is challenging to discern the precise coupling at the atomic scale, given the unclear dependence of exchange bias on temperature or gate voltage. The interfacial coupling likely represents an averaged effect, with atomic details at the interface being less distinct. As a future direction, we plan to investigate the interface size dependence on exchange bias, aiming for a clearer understanding.

It could also explain why approximately 43\% of the magnetization was altered, as the transport measurements reflect an average effect from both the tuned surface layers and the intrinsic FGaT layers.
 


\section{Conclusion}
With the realization of room-temperature multiferroic control of magnetism, this work holds promise for the future development of ultracompact and energy-efficient spintronic devices. Additionally, exchange bias observed in this work and others is compared and listed in Table \ref{Tab-exchange}. Further enhancement of exchange bias can be achieved by reducing the ferromagnet thickness and utilizing ferroelectric control of the ferromagnet/antiferromagnet interface.
Additionally, the multiple resistance states highlight promising applications in memristors for neuromorphic computing, introducing a new functional capability for 2D vdW magnet-based devices. Moreover, magnetic skyrmions, which feature real-space topology \cite{wang2020topological,pan2022efficient}, are emerging as candidates for future neuromorphic devices. Skyrmions have been demonstrated in FGaT and Fe$_3$GeTe$_2$-based structures \cite{wu2020neel,wu2022van,liu2024magnetic}. The integration of ferroelectric control could further enable ultra energy-efficient skyrmion manipulation, unlocking new opportunities for next-generation spintronic technologies.

\begin{table}
\centering
\caption{Exchange bias in vdW systems.}\label{Tab-exchange}
\begin{tabular}{|c|c|c|c|}\hline
Structure & FM thickness & Exchange bias & Temperature \\ \hline
CGT/FGeT & 4 nm & $\sim$ 5 mT\cite{wu2022manipulating} & 2 K \\ \hline
\textbf{CIPS/FGaT} & 20 nm & \makecell{13.8 mT \\(\textbf{this work})} & 5 K\\ \hline
FGeT/MnPS$_3$ & 23 nm \cite{dai2021enhancement}& 15 mT & 10 K \\\hline
FGeT/MnPSe$_3$ & 23 nm \cite{dai2021enhancement} & 22 mT & 10 K \\ \hline
FGeT/O-FGeT & $>$100 nm & 35 mT\cite{gweon2021exchange} & 70 K \\ \hline
CrCl$_3$/FGeT & 30 nm & 56 mT\cite{zhu2020exchange} & 2.5 K \\ \hline
FGeT/FePS$_3$& 18 nm  & 60 mT\cite{huang2023manipulating} & 20 K 12 Gpa \\ \hline
O-FGeT/FGeT/CrSe & 16 nm & 90 mT\cite{wu2022manipulating} & 5 K \\ \hline
\multicolumn{4}{|c|}{FM=ferromagnet, CGT=Cr$_2$Ge$_2$Te$_6$ FGeT=Fe$_3$GeTe$_2$} \\ \hline
\end{tabular}
\vspace{5pt}
\end{table}


\appendices


\section*{Acknowledgment}

Y. Wu is supported by University of Florida start-up funds and 2024 Research Opportunity Seed Fund (ROSF). Z.Sofer was supported by ERC-CZ program (project LL2101) from Ministry of Education Youth and Sports (MEYS) and by the project Advanced Functional Nanorobots (reg. No. CZ.02.1.01/0.0/0.0/15\_003/0000444 financed by the EFRR).

\ifCLASSOPTIONcaptionsoff
  \newpage
\fi

\bibliographystyle{IEEEtran}
\bibliography{FE-FM.bib}

\begin{thebibliography}{10}
\providecommand{\url}[1]{#1}
\csname url@samestyle\endcsname
\providecommand{\newblock}{\relax}
\providecommand{\bibinfo}[2]{#2}
\providecommand{\BIBentrySTDinterwordspacing}{\spaceskip=0pt\relax}
\providecommand{\BIBentryALTinterwordstretchfactor}{4}
\providecommand{\BIBentryALTinterwordspacing}{\spaceskip=\fontdimen2\font plus
\BIBentryALTinterwordstretchfactor\fontdimen3\font minus \fontdimen4\font\relax}
\providecommand{\BIBforeignlanguage}[2]{{%
\expandafter\ifx\csname l@#1\endcsname\relax
\typeout{** WARNING: IEEEtran.bst: No hyphenation pattern has been}%
\typeout{** loaded for the language `#1'. Using the pattern for}%
\typeout{** the default language instead.}%
\else
\language=\csname l@#1\endcsname
\fi
#2}}
\providecommand{\BIBdecl}{\relax}
\BIBdecl

\bibitem{zhang20232d}
B.~Zhang, P.~Lu, R.~Tabrizian, P.~X.-L. Feng, and Y.~Wu, ``2{D} magnetic heterostructures: {S}pintronics and quantum future,'' \emph{npj Spintronics}, vol.~2, no.~6, 2024.

\bibitem{hu2019perspective}
J.-M. Hu, C.-W. Nan, and L.-Q. Chen, ``Perspective: voltage control of magnetization in multiferroic heterostructures,'' \emph{National Science Review}, vol.~6, no.~4, pp. 621--624, 2019.

\bibitem{ramesh2021electric}
R.~Ramesh and S.~Manipatruni, ``Electric field control of magnetism,'' \emph{Proceedings of the Royal Society A}, vol. 477, no. 2251, p. 20200942, 2021.

\bibitem{chen2016probing}
X.~Chen, L.~Wang, Y.~Wu, H.~Gao, Y.~Wu, G.~Qin, Z.~Wu, Y.~Han, S.~Xu, T.~Han, W.~Ye, J.~Lin, G.~Long, Y.~He, Y.~Cai, W.~Ren, and N.~Wang, ``Probing the electronic states and impurity effects in black phosphorus vertical heterostructures,'' \emph{2D Materials}, vol.~3, no.~1, p. 015012, 2016.

\bibitem{wu2016negative}
Y.~Wu, X.~Chen, Z.~Wu, S.~Xu, T.~Han, J.~Lin, B.~Skinner, Y.~Cai, Y.~He, C.~Cheng, and N.~Wang, ``Negative compressibility in graphene-terminated black phosphorus heterostructures,'' \emph{Physical Review B}, vol.~93, no.~3, p. 035455, 2016.

\bibitem{zhong2024integrating}
H.~Zhong, P.~Plummer, Douglas Z.and~Lu, Y.~Li, P.~A. Leger, and Y.~Wu, ``Integrating 2{D} magnets for quantum devices: from materials and characterization to future technology,'' \emph{arXiv preprint arXiv:2406.12136}, 2024.

\bibitem{hendriks2024electric}
F.~Hendriks, R.~R. Rojas-Lopez, B.~Koopmans, and M.~H. Guimar{\~a}es, ``Electric control of optically-induced magnetization dynamics in a van der {W}aals ferromagnetic semiconductor,'' \emph{Nature Communications}, vol.~15, no.~1, p. 1298, 2024.

\bibitem{wang2024manipulation}
R.-Q. Wang, T.-M. Lei, and Y.-W. Fang, ``Manipulation of magnetic anisotropy of 2{D } magnetized graphene by ferroelectric {I}n$_2${S}e$_3$,'' \emph{Journal of Applied Physics}, vol. 135, no.~8, 2024.

\bibitem{min2023reduced}
S.~Min, R.~Wang, Y.~Wang, K.~Song, and Z.~Chu, ``Reduced resonance frequency and enhanced coupling coefficient in fishtailing magnetoelectric resonator,'' \emph{Applied Physics Letters}, vol. 123, no.~15, 2023.

\bibitem{chen2019giant}
A.~Chen, Y.~Wen, B.~Fang, Y.~Zhao, Q.~Zhang, Y.~Chang, P.~Li, H.~Wu, H.~Huang, Y.~Lu \emph{et~al.}, ``Giant nonvolatile manipulation of magnetoresistance in magnetic tunnel junctions by electric fields via magnetoelectric coupling,'' \emph{Nature Communications}, vol.~10, no.~1, p. 243, 2019.

\bibitem{borisov2005magnetoelectric}
P.~Borisov, A.~Hochstrat, X.~Chen, W.~Kleemann, and C.~Binek, ``Magnetoelectric switching of exchange bias,'' \emph{Physical Review Letters}, vol.~94, no.~11, p. 117203, 2005.

\bibitem{leger2024machine}
P.~A. Leger, A.~Ramesh, T.~Ulloa, and Y.~Wu, ``Machine-learning-enabled fast optical identification and characterization of 2{D} materials,'' \emph{arXiv preprint arXiv:2406.16211}, 2024.

\bibitem{han2018investigation}
T.~Han, J.~Shen, N.~F. Yuan, J.~Lin, Z.~Wu, Y.~Wu, S.~Xu, L.~An, G.~Long, Y.~Wang \emph{et~al.}, ``Investigation of the two-gap superconductivity in a few-layer {N}b{S}e$_2$-graphene heterojunction,'' \emph{Physical Review B}, vol.~97, no.~6, p. 060505, 2018.

\bibitem{wu2019induced}
Y.~Wu, J.~J. He, T.~Han, S.~Xu, Z.~Wu, J.~Lin, T.~Zhang, Y.~He, and N.~Wang, ``Induced ising spin-orbit interaction in metallic thin films on monolayer {WS}e$_2$,'' \emph{Physical Review B}, vol.~99, no.~12, p. 121406, 2019.

\bibitem{wu2020neel}
Y.~Wu, S.~Zhang, J.~Zhang, W.~Wang, Y.~L. Zhu, J.~Hu, G.~Yin, K.~Wong, C.~Fang, C.~Wan \emph{et~al.}, ``N{\'e}el-type skyrmion in {WT}e$_2$/{F}e$_3${G}e{T}e$_2$ van der {W}aals heterostructure,'' \emph{Nature Communications}, vol.~11, no.~1, p. 3860, 2020.

\bibitem{velicky2018mechanism}
M.~Velicky, G.~E. Donnelly, W.~R. Hendren, S.~McFarland, D.~Scullion, W.~J. DeBenedetti, G.~C. Correa, Y.~Han, A.~J. Wain, M.~A. Hines, D.~A. Muller, K.~S. Novoselov, H.~D. Abruna, R.~M. Bowman, E.~J.~G. Santos, and F.~Huang, ``Mechanism of gold-assisted exfoliation of centimeter-sized transition-metal dichalcogenide monolayers,'' \emph{ACS Nano}, vol.~12, no.~10, pp. 10\,463--10\,472, 2018.

\bibitem{huang2020universal}
Y.~Huang, Y.-H. Pan, R.~Yang, L.-H. Bao, L.~Meng, H.-L. Luo, Y.-Q. Cai, G.-D. Liu, W.-J. Zhao, Z.~Zhou \emph{et~al.}, ``Universal mechanical exfoliation of large-area 2{D} crystals,'' \emph{Nature Communications}, vol.~11, no.~1, p. 2453, 2020.

\bibitem{huang2017layer}
B.~Huang, G.~Clark, E.~Navarro-Moratalla, D.~R. Klein, R.~Cheng, K.~L. Seyler, D.~Zhong, E.~Schmidgall, M.~A. McGuire, D.~H. Cobden \emph{et~al.}, ``Layer-dependent ferromagnetism in a van der {W}aals crystal down to the monolayer limit,'' \emph{Nature}, vol. 546, no. 7657, pp. 270--273, 2017.

\bibitem{huang2018electrical}
B.~Huang, G.~Clark, D.~R. Klein, D.~MacNeill, E.~Navarro-Moratalla, K.~L. Seyler, N.~Wilson, M.~A. McGuire, D.~H. Cobden, D.~Xiao \emph{et~al.}, ``Electrical control of 2{D} magnetism in bilayer {C}r{I}$_3$,'' \emph{Nature Nanotechnology}, vol.~13, no.~7, pp. 544--548, 2018.

\bibitem{wang2018electric}
Z.~Wang, T.~Zhang, M.~Ding, B.~Dong, Y.~Li, M.~Chen, X.~Li, J.~Huang, H.~Wang, X.~Zhao \emph{et~al.}, ``Electric-field control of magnetism in a few-layered van der {W}aals ferromagnetic semiconductor,'' \emph{Nature nanotechnology}, vol.~13, no.~7, pp. 554--559, 2018.

\bibitem{deng2018gate}
Y.~Deng, Y.~Yu, Y.~Song, J.~Zhang, N.~Z. Wang, Z.~Sun, Y.~Yi, Y.~Z. Wu, S.~Wu, J.~Zhu \emph{et~al.}, ``Gate-tunable room-temperature ferromagnetism in two-dimensional {F}e$_3${G}e{T}e$_2$,'' \emph{Nature}, vol. 563, no. 7729, pp. 94--99, 2018.

\bibitem{toprasertpong2022strong}
K.~Toprasertpong, M.~Takenaka, and S.~Takagi, ``On the strong coupling of polarization and charge trapping in {H}f{O}$_2$/{S}i-based ferroelectric field-effect transistors: overview of device operation and reliability,'' \emph{Applied Physics A}, vol. 128, no.~12, p. 1114, 2022.

\bibitem{zhang2010tuning}
J.~Zhang, C.~Yang, S.~Wu, Y.~Liu, M.~Zhang, H.~Chen, W.~Zhang, and Y.~Li, ``Tuning two-dimensional electron gas of ferroelectric/{G}a{N} heterostructures by ferroelectric polarization,'' \emph{Semiconductor Science and Technology}, vol.~25, no.~3, p. 035011, 2010.

\bibitem{zhang2022above}
G.~Zhang, F.~Guo, H.~Wu, X.~Wen, L.~Yang, W.~Jin, W.~Zhang, and H.~Chang, ``Above-room-temperature strong intrinsic ferromagnetism in 2{D} van der {W}aals {F}e$_3${G}a{T}e$_2$ with large perpendicular magnetic anisotropy,'' \emph{Nature Communications}, vol.~13, no.~1, p. 5067, 2022.

\bibitem{ma2020high}
R.-R. Ma, D.-D. Xu, Z.~Guan, X.~Deng, F.~Yue, R.~Huang, Y.~Chen, N.~Zhong, P.-H. Xiang, and C.-G. Duan, ``High-speed ultraviolet photodetectors based on 2{D} layered {C}u{I}n{P}$_2${S}$_6$ nanoflakes,'' \emph{Applied Physics Letters}, vol. 117, no.~13, 2020.

\bibitem{chen2021reversal}
Z.~Chen, Q.~Yang, L.~Tao, and E.~Y. Tsymbal, ``Reversal of the magnetoelectric effect at a ferromagnetic metal/ferroelectric interface induced by metal oxidation,'' \emph{npj Computational Materials}, vol.~7, no.~1, p. 204, 2021.

\bibitem{song2017recent}
C.~Song, B.~Cui, F.~Li, X.~Zhou, and F.~Pan, ``Recent progress in voltage control of magnetism: {M}aterials, mechanisms, and performance,'' \emph{Progress in Materials Science}, vol.~87, pp. 33--82, 2017.

\bibitem{gong2018electrically}
S.-J. Gong, C.~Gong, Y.-Y. Sun, W.-Y. Tong, C.-G. Duan, J.-H. Chu, and X.~Zhang, ``Electrically induced 2{D} half-metallic antiferromagnets and spin field effect transistors,'' \emph{Proceedings of the National Academy of Sciences}, vol. 115, no.~34, pp. 8511--8516, 2018.

\bibitem{schlom2007strain}
D.~G. Schlom, L.-Q. Chen, C.-B. Eom, K.~M. Rabe, S.~K. Streiffer, and J.-M. Triscone, ``Strain tuning of ferroelectric thin films,'' \emph{Annual Review of Materials Research}, vol.~37, no.~1, pp. 589--626, 2007.

\bibitem{fang2003anomalous}
Z.~Fang, N.~Nagaosa, K.~S. Takahashi, A.~Asamitsu, R.~Mathieu, T.~Ogasawara, H.~Yamada, M.~Kawasaki, Y.~Tokura, and K.~Terakura, ``The anomalous {H}all effect and magnetic monopoles in momentum space,'' \emph{Science}, vol. 302, no. 5642, pp. 92--95, 2003.

\bibitem{zheng2021berry}
D.~Zheng, Y.-W. Fang, S.~Zhang, P.~Li, Y.~Wen, B.~Fang, X.~He, Y.~Li, C.~Zhang, W.~Tong \emph{et~al.}, ``Berry phase engineering in {S}r{R}u{O}$_3$/{S}r{I}r{O}$_3$/{S}r{T}i{O}$_3$ superlattices induced by band structure reconstruction,'' \emph{ACS Nano}, vol.~15, no.~3, pp. 5086--5095, 2021.

\bibitem{nagaosa2010anomalous}
N.~Nagaosa, J.~Sinova, S.~Onoda, A.~H. MacDonald, and N.~P. Ong, ``Anomalous {H}all effect,'' \emph{Reviews of Modern Physics}, vol.~82, no.~2, p. 1539, 2010.

\bibitem{gweon2021exchange}
H.~K. Gweon, S.~Y. Lee, H.~Y. Kwon, J.~Jeong, H.~J. Chang, K.-W. Kim, Z.~Q. Qiu, H.~Ryu, C.~Jang, and J.~W. Choi, ``Exchange bias in weakly interlayer-coupled van der {W}aals magnet {F}e$_3${G}e{T}e$_2$,'' \emph{Nano letters}, vol.~21, no.~4, pp. 1672--1678, 2021.

\bibitem{wu2022manipulating}
Y.~Wu, W.~Wang, L.~Pan, and K.~L. Wang, ``Manipulating exchange bias in a van der {W}aals ferromagnet,'' \emph{Advanced Materials}, vol.~34, no.~12, p. 2105266, 2022.

\bibitem{huang2023manipulating}
X.~Huang, L.~Zhang, L.~Tong, Z.~Li, Z.~Peng, R.~Lin, W.~Shi, K.-H. Xue, H.~Dai, H.~Cheng, D.~d.~C. Branco, J.~Xu, J.~Han, G.~J. Cheng, X.~Miao, and L.~Ye, ``Manipulating exchange bias in 2{D} magnetic heterojunction for high-performance robust memory applications,'' \emph{Nature Communications}, vol.~14, no.~1, p. 2190, 2023.

\bibitem{wu2022van}
Y.~Wu, B.~Francisco, Z.~Chen, W.~Wang, Y.~Zhang, C.~Wan, X.~Han, H.~Chi, Y.~Hou, A.~Lodesani \emph{et~al.}, ``A van der {W}aals interface hosting two groups of magnetic skyrmions,'' \emph{Advanced Materials}, vol.~34, no.~16, p. 2110583, 2022.

\bibitem{wang2020topological}
K.~L. Wang, Y.~Wu, C.~Eckberg, G.~Yin, and Q.~Pan, ``Topological quantum materials,'' \emph{MRS Bulletin}, vol.~45, no.~5, pp. 373--379, 2020.

\bibitem{pan2022efficient}
Q.~Pan, Y.~Liu, H.~Wu, P.~Zhang, H.~Huang, C.~Eckberg, X.~Che, Y.~Wu, B.~Dai, Q.~Shao \emph{et~al.}, ``Efficient spin-orbit torque switching of perpendicular magnetization using topological insulators with high thermal tolerance,'' \emph{Advanced Electronic Materials}, vol.~8, no.~9, p. 2200003, 2022.

\bibitem{liu2024magnetic}
C.~Liu, S.~Zhang, H.~Hao, H.~Algaidi, Y.~Ma, and X.-X. Zhang, ``Magnetic skyrmions above room temperature in a van der {W}aals ferromagnet {F}e$_3${G}a{T}e$_2$,'' \emph{Advanced Materials}, p. 2311022, 2024.

\bibitem{dai2021enhancement}
H.~Dai, H.~Cheng, M.~Cai, Q.~Hao, Y.~Xing, H.~Chen, X.~Chen, X.~Wang, and J.-B. Han, ``Enhancement of the coercive field and exchange bias effect in {F}e$_3${G}e{T}e$_2$/{M}n{PX}$_3$ ({X}= {S} and {S}e) van der {W}aals heterostructures,'' \emph{ACS Applied Materials \& Interfaces}, vol.~13, no.~20, pp. 24\,314--24\,320, 2021.

\bibitem{zhu2020exchange}
R.~Zhu, W.~Zhang, W.~Shen, P.~K.~J. Wong, Q.~Wang, Q.~Liang, Z.~Tian, Y.~Zhai, C.-w. Qiu, and A.~T. Wee, ``Exchange bias in van der {W}aals {C}r{Cl}$_3$/{F}e$_3${G}e{T}e$_2$ heterostructures,'' \emph{Nano Letters}, vol.~20, no.~7, pp. 5030--5035, 2020.

\end{thebibliography}

\end{document}